\documentclass[sigconf]{acmart}
\usepackage[vlined, ruled, linesnumbered]{algorithm2e}
\usepackage{amssymb}
\usepackage{amsfonts}
\usepackage{amsmath}

\AtBeginDocument{%
  \providecommand\BibTeX{{%
    \normalfont B\kern-0.5em{\scshape i\kern-0.25em b}\kern-0.8em\TeX}}}

\setcopyright{none}
\renewcommand\footnotetextcopyrightpermission[1]{}
\acmConference[Woodstock '18]{Woodstock '18: ACM Symposium on Neural
	Gaze Detection}{June 03--05, 2018}{Woodstock, NY}
\acmBooktitle{Woodstock '18: ACM Symposium on Neural Gaze Detection,
	June 03--05, 2018, Woodstock, NY}
\acmPrice{15.00}
\acmISBN{978-1-4503-9999-9/18/06}

\acmSubmissionID{182}

\begin{document}

%%
%% The "title" command has an optional parameter,
%% allowing the author to define a "short title" to be used in page headers.

\title{SocialTrans: A Deep Sequential Model with Social Information for Web-Scale Recommendation Systems}

%%
%% The "author" command and its associated commands are used to define
%% the authors and their affiliations.
%% Of note is the shared affiliation of the first two authors, and the
%% "authornote" and "authornotemark" commands
%% used to denote shared contribution to the research.
\author{Qiaoan Chen}
\author{Hao Gu}
\author{Lingling Yi}
\affiliation{\institution{Weixin Group, Tencent Inc.}}
\email{{kazechen, nickgu, chrisyi}@tencent.com}

\author{Yishi Lin}
\author{Peng He}
\author{Chuan Chen}
\affiliation{\institution{Weixin Group, Tencent Inc.}}
\email{{elsielin, paulhe, chuanchen}@tencent.com}

\author{Yangqiu Song}
\affiliation{\institution{Department of CSE, Hong Kong University of Science and Technology}}
\email{yqsong@cse.ust.hk}

%%
%% By default, the full list of authors will be used in the page
%% headers. Often, this list is too long, and will overlap
%% other information printed in the page headers. This command allows
%% the author to define a more concise list
%% of authors' names for this purpose.
%% \renewcommand{\shortauthors}{Trovato and Tobin, et al.}

%%
%% The abstract is a short summary of the work to be presented in the
%% article.
\begin{abstract}
On social network platforms, a user's behavior is based on his/her personal interests, or influenced by his/her friends. In the literature, it is common to model either users' personal preference or their socially influenced preference. In this paper, we present a novel deep learning model SocialTrans for social recommendations to integrate these two types of preferences. SocialTrans is composed of three modules. The first module is based on a multi-layer Transformer to model users' personal preference. The second module is a multi-layer graph attention neural network (GAT), which is used to model the social influence strengths between friends in social networks. The last module merges users' personal preference and socially influenced preference to produce recommendations. Our model can efficiently fit large-scale data and we deployed SocialTrans to a major article recommendation system in China. Experiments on three data sets verify the effectiveness of our model and show that it outperforms state-of-the-art social recommendation methods.
\end{abstract}

%%
%% The code below is generated by the tool at http://dl.acm.org/ccs.cfm.
%% Please copy and paste the code instead of the example below.
%%
%%\begin{CCSXML}
%%<ccs2012>
%% <concept>
%%  <concept_id>10010520.10010553.10010562</concept_id>
%%  <concept_desc>Computer systems organization~Embedded systems</concept_desc>
%%  <concept_significance>500</concept_significance>
%% </concept>
%% <concept>
%%  <concept_id>10010520.10010575.10010755</concept_id>
%%  <concept_desc>Computer systems organization~Redundancy</concept_desc>
%%  <concept_significance>300</concept_significance>
%% </concept>
%% <concept>
%%  <concept_id>10010520.10010553.10010554</concept_id>
%%  <concept_desc>Computer systems organization~Robotics</concept_desc>
%%  <concept_significance>100</concept_significance>
%% </concept>
%% <concept>
%%  <concept_id>10003033.10003083.10003095</concept_id>
%%  <concept_desc>Networks~Network reliability</concept_desc>
%%  <concept_significance>100</concept_significance>
%% </concept>
%%</ccs2012>
%%\end{CCSXML}

%%\ccsdesc[500]{Computer systems organization~Embedded systems}
%%\ccsdesc[300]{Computer systems organization~Redundancy}
%%\ccsdesc{Computer systems organization~Robotics}
%%\ccsdesc[100]{Networks~Network reliability}

%%
%% Keywords. The author (s) should pick words that accurately describe
%% the work being presented. Separate the keywords with commas.
%%\keywords{Social Recommendation, Graph Attention Neural Networks, Transformer, Social Influence}

%%
%% This command processes the author and affiliation and title
%% information and builds the first part of the formatted document.
\maketitle

\section{INTRODUCTION} \label{section:introduction}
Social network platforms such as Facebook and Twitter are very popular and they have become an essential part of our daily life. These platforms provide places for people to communicate with each other. On these platforms, users can share information (e.g., articles, videos and games) with their friends. To enrich user experiences, these platforms often build recommendation systems to help their users to explore new things, for example by listing "things you may be interested in". Recommendation systems deployed in social network platforms usually use users' profiles and their history behaviors to make predictions about their interests. In social network platforms, users' behavior could also be significantly influenced by their friends. Thus, it is crucial to incorporate social influence in the recommendation systems, which motivates this work.

\begin{figure}[h]
	\centering
	\includegraphics[width=\linewidth]{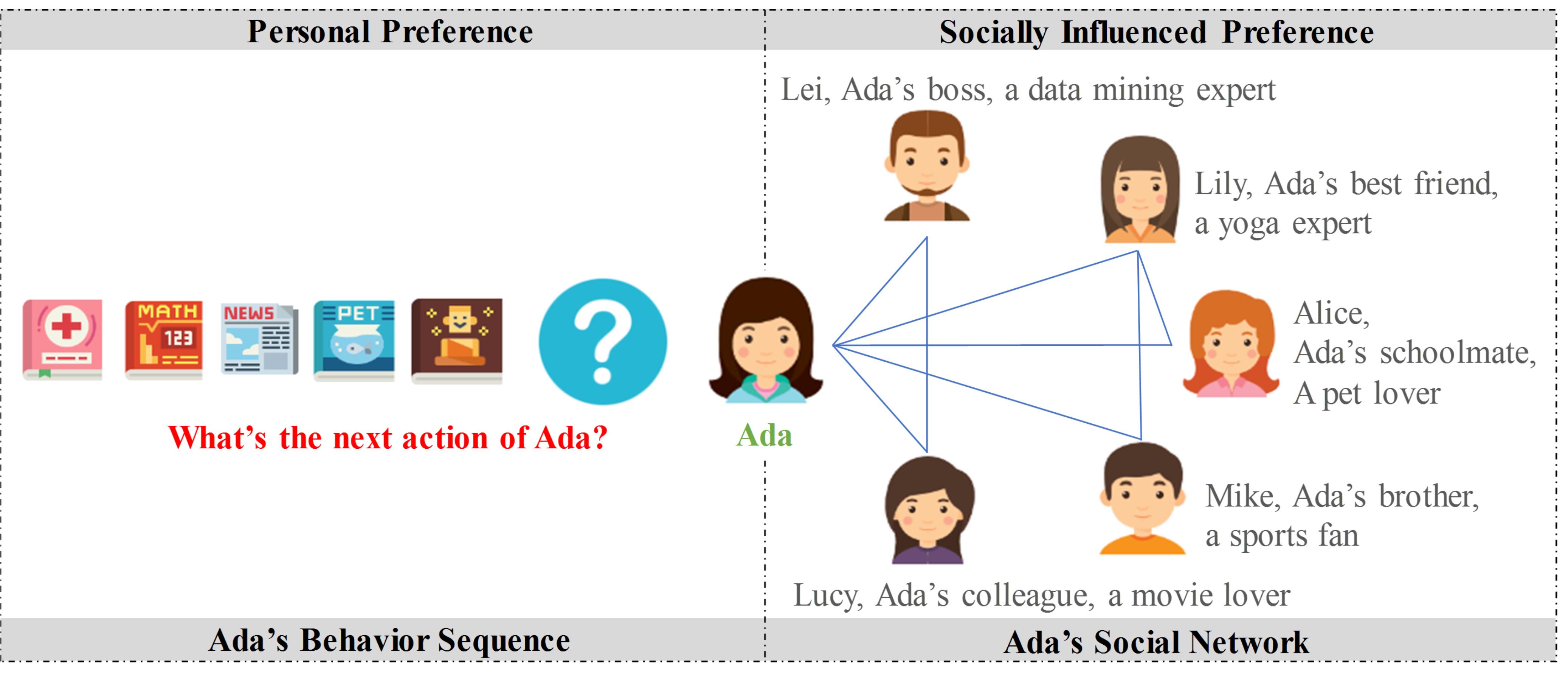}
	\caption{An illustration of Ada's historical behavior and her social network.}
	\label{figure:introduction_example}
\end{figure}

Figure \ref{figure:introduction_example} presents how Ada behaves in an online community.\footnote{Icons made by Freepik from \url{www.flaticon.com}.} The left part is her historical behavior, described by a sequence of actions (e.g., item clicks), and the right part is her social network. First, user interests are dynamic by nature. Ada has been interested in pets for a long period, but she may search for yoga books in the future. We should capture Ada's dynamic interest from her behaviors. Second, Ada trusts her boss who is an expert in data mining when searching for technology news, while she could be influenced by another friend when searching for yoga. This socially influenced preference should be considered in modeling.

To get deeper insights into this phenomenon, we analyze a real-world social network platform - WeChat, a popular mobile application in China, with more than one billion monthly active users\footnote{\url{https://www.wechat.com/en}}. WeChat users can read and share articles with their friends. In this analysis, if a user shares an article and his friend (or an $n$-hop friend) re-shares it, we say that his friend is influenced by him. Let $H(n)$ be the average influence probability for each user and his $n$-hop friend pairs, and $H(0)$ be the average sharing probability. This analysis answers two questions: (1) how social influence strength changes in different hops; (2) how social influence strength varies in different topics.

\begin{figure}[h]
	\centering
	\includegraphics[width=\linewidth]{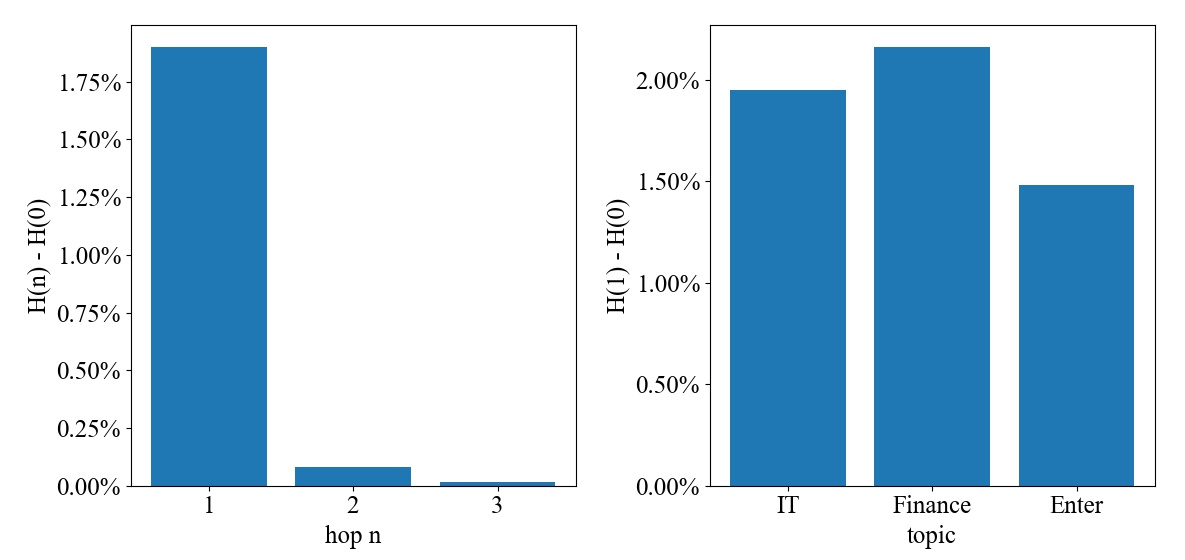}
	\caption{Analysis of social influence in WeChat. $H(n)$ represents the average influence probability for each user and his $n$-hop friend pairs, while $H(0)$ represents the average sharing probability. All users are anonymous in this analysis. "Enter" is the abbreviation for entertainment. }
	\label{figure:introduction_analysis}
\end{figure}

Figure \ref{figure:introduction_analysis} shows the analysis result. In the left part, we consider the increased probability of influence strength $H(n) - H(0)$, which describes how significantly a user is influenced by his n-hop friends compared to a global probability. It shows that users are significantly influenced by 1-hop friends and the influence strength decreases dramatically when the hop increases. The right part of the Figure \ref{figure:introduction_analysis} shows that direct friends' influence $H(1)$ is quite different in various topics. These results motivate us to model context-dependent social influence to improve the recommendation system.

In this paper, we propose an approach to model users' personal preferences and context-dependent socially influenced preferences. Our recommendation model, named SocialTrans, is based on two recent state-of-the-art models, Transformer \cite{DBLP:conf/nips/VaswaniSPUJGKP17} and graph-attention network (GAT) \cite{velivckovic2017graph}. A multi-layer Transformer is used to capture users' personal preferences. Socially influenced preferences are captured by a multi-layer GAT, extended by considering edge attributes and the multi-head attention mechanism. We conduct offline experiments on two data sets and online A/B testing to verify our model. The results show that SocialTrans achieves the state-of-the-art performance and achieves at least a $9.5\%$ relative improvement on offline data sets and a $5.89\%$ relative improvement on online A/B testing. Our contributions can be summarized as follows:
\begin{itemize}
	\item \textbf{Novel methodologies}. We propose SocialTrans to model both users' personal preferences and their socially influenced preferences. We combine Transformer and GAT for social recommendation tasks. In particular, the GAT we use is an extended version that considers edge attributes and uses the multi-head attention mechanism to model context-dependent social influence.
	\item \textbf{Multifaceted experiments}. We evaluate our approach on one benchmark data set and two commercial data sets. The experimental results verify the superiority of our approach over state-of-the-art techniques. 
	\item \textbf{Large-scale implementation}. We train and deploy SocialTrans in a real-world recommendation system which can potentially affects over one billion users. This model contains a three-layer Transformer and a two-layer GAT. We provide techniques to speed up both offline training and online service procedures.
	\item \textbf{Economical online evaluation procedure}. Model evaluation in a fast-growing recommendation system is computationally expensive. Many items can be added or fading-away every day. We provide an efficient deployment and evaluation procedure to overcome the difficulties.
\end{itemize}
\textbf{Organization}. We first formulate the problem in Section \ref{section:definition}. Then, we introduce our proposed model SocialTrans in Section \ref{section:model_framework}. Large scale implementation details are in Section \ref{section:large_scale}. Section \ref{section:experiments} shows our experimental results. Related works are in Section \ref{section:realted_work}. Section \ref{section:conclusion} concludes the paper.

\section{Problem Definition} \label{section:definition}
The goal of sequence-based social recommendation is to predict which item a user will click soon, based on his previous clicked history and the social network. In this setting, let $U$ denote the set of users and $V$ be the set of items. We use $G = (U, E)$ to denote the social network, where $E$ is the set of friendship links between users. At each timestamp $t$, user $u$'s previous behavior sequence is represented by an ordered list $S_{t-1}^{u} =\left(v_{0}^{u}, v_{1}^{u}, v_{2}^{u} , \cdots, v_{t-1}^{u}\right)$, $u \in U, v^u_j \in V, 1 \leq j \leq t-1$. Sequence-based social recommendation utilizes both information from a user $u$ and his friends, which can be represented as $\mathbb{S}^{u}_{t-1} = \{S_{t-1}^{u'} |  u' \in \{u\} \cup N(u)\}$. Here $N(u)$ is the set of $u$'s friends. Given $\mathbb{S}^{u}_{t-1}$, sequence-based social recommendation aims to predict which item $v$ is likely to be clicked by user $u$.

In a real-world data set, the length of a user's behavior sequence can be up to several hundreds, which is hard for many models to handle. To simplify our question, we transform a user's previous behavior sequence $S_{t-1}^{u}=\left(v_{0}^{u}, v_{1}^{u}, v_{2}^{u} , \cdots, v_{t-1}^{u}\right)$ into a fixed length sequence $\hat{S}_{t-1}^{u} = (v_0, v_1, ..., v_{m-1}), v_j \in V, 0 \leq j \leq m-1$. Here $m$ represents the maximum length that our model can handle and $\hat{S}_{t-1}^{u}$ is most recent $m$ items in $S_{t-1}^{u}$. If the sequence length is less than $m$, we repeatedly add a ‘blank’ item to the left until the length is $m$. Similarly, $\hat{\mathbb{S}}^{u}_{t-1}$ represents the fixed-length version of sequences in $\mathbb{S}^{u}_{t-1}$. How to handle longer length of sequence will be left out as future work.

\section{Model Framework} \label{section:model_framework}
\begin{figure*}
	\centering
	\includegraphics[width=\textwidth]{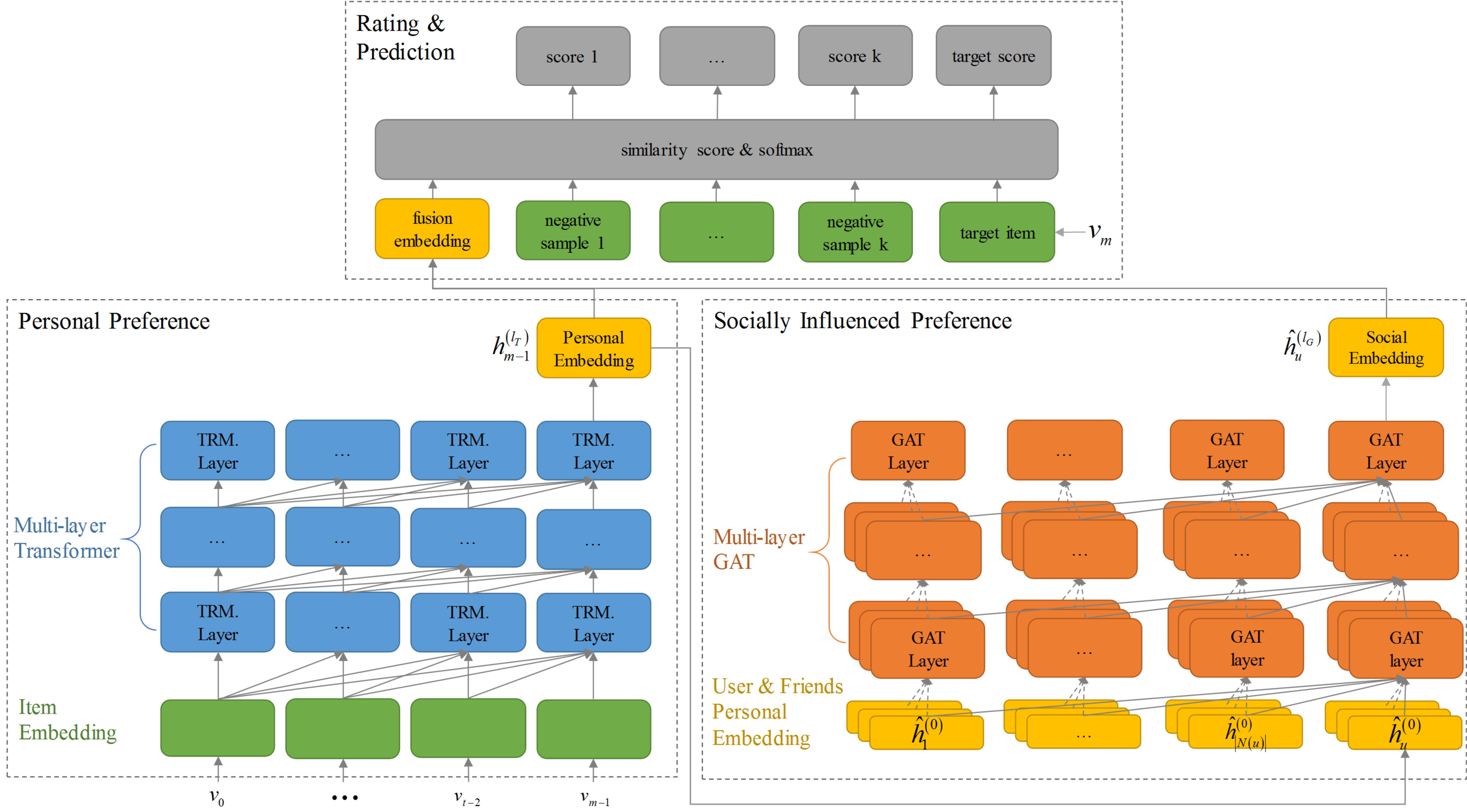}
	\caption{SocialTrans model architecture. It contains three modules: a multi-layer Transformer for personal preference modeling, a multi-layer GAT for socially influenced modeling and a rating \& prediction module.}
	\label{figure:model_framework}
\end{figure*}

Motivated by the observation that a user’s behavior can be determined by personal preference and socially influenced preference, we propose a novel method named SocialTrans for recommendation systems in social-network platforms. Figure \ref{figure:model_framework} provides an overview of SocialTrans. SocialTrans is composed of three modules: personal preference modeling, socially influenced preference modeling, and rating prediction. First, a user's personal preference is modeled by a multi-layer Transformer \cite{DBLP:conf/icdm/KangM18}, which can capture his dynamic interest (§\ref{subsection:self_preference}). Second, a multi-layer GAT \cite{velivckovic2017graph} is used to model socially influenced preference from his friends. The GAT we use is an extended version that considers edge attributes and uses the multi-head attention mechanism (§\ref{subsection:social_influence_preference}). Finally, the user's personal preference and socially influenced preference are fused to get the final representation. Rating scores between users and items are computed to produce recommendation results (§\ref{subsection:recommendation}).

\subsection{Personal Preference Modeling} \label{subsection:self_preference}
The personal preference modeling module tries to capture how users' dynamic interests evolve over time. To be specific, this module generates a user's personal preference embedding at the current timestamp given his behavior sequence.

\begin{figure}[h]
	\centering
	\includegraphics[width=0.4\linewidth]{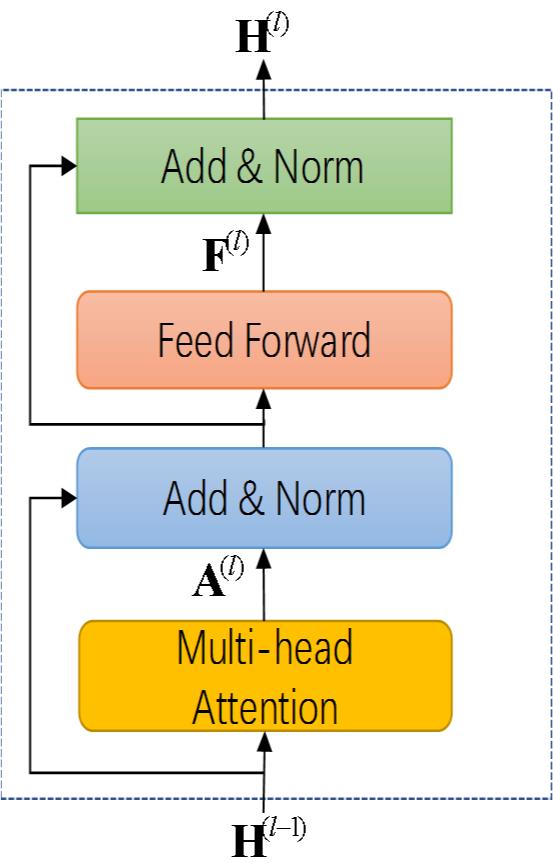}
	\caption{Illustration of a layer in Transformer. }
	\label{figure:transformer}
\end{figure}

We use a multi-layer Transformer \cite{DBLP:conf/icdm/KangM18} to capture users' personal preferences. Transformer is widely used in sequence modeling. It is able to capture the correlation between any pairs in sequences. As shown in Figure \ref{figure:transformer}, the Transformer layer contains three sub-layers, a Multi-Head Attention sub-layer, a Feed-Forward Network sub-layer, and an Add \& Norm sub-layer. We now describe the input, the output, and sub-layers in Transformer in detail.

{\bfseries Input Embedding}. The input matrix $\mathbf{H}^{(0)} \in \mathbb{R}^{m \times d}$ to the multi-layer Transformer is mainly constructed from items in user's behavior sequence $\hat{S}_{t-1}^{u} = (v_0, v_1, ..., v_{m-1})$. Here $d$ is the hidden dimension and each item $v \in V$ is represented as a row $\mathbf{w}_v$ in the item embedding matrix $\mathbf{W} \in \mathbb{R}^{|V| \times d}$. Since the Transformer can't be aware of items' position, each position $\tau$ is associated with a learnable position embedding vector $\mathbf{p}_\tau \in \mathbb{R}^d$ to carry location information for the corresponding item. Each row $\mathbf{h}^{(0)}_\tau$ in $\mathbf{H}^{(0)}$ is defined as:
\begin{equation}
\mathbf{h}^{(0)}_\tau = \mathbf{w}_{v_\tau} + \mathbf{p}_\tau
\end{equation}

{\bfseries Multi-head Self-Attention}. Attention mechanisms are widely used in sequence modeling tasks. They allow a model to capture the relationship between any pairs in the sequences. Recent work \cite{li2018multi, devlin2019bert} has shown that attention to different representation subspaces simultaneously is beneficial. In this work, we adopt the multi-head self-attention as in work \cite{DBLP:conf/nips/VaswaniSPUJGKP17}. This allows the model to jointly attend to information from different representation subspaces. 

First, the scaled dot-product attention is applied to each head. This attention function can be described as mapping a set of query-key-value tuples to an output. It is defined as:
\begin{equation}
\text{Attention}(\mathbf{Q}, \mathbf{K}, \mathbf{V}) = \text{softmax}(\frac{\mathbf{Q} \mathbf{K}^T} {\sqrt{d_s}}) \mathbf{V}
\end{equation}
Here $\mathbf{Q}, \mathbf{K}, \mathbf{V}  \in \mathbb{R}^{m \times d_s}$ represent the query, key, and value matrices. Moreover, $d_s$ is the dimensionality for each head, and we have $d_s = d / r$ for an $r$-head model. The scaled dot-product attention computes a weighted sum of all values, where the weight is calculated by the query matrix $\mathbf{Q}$ and the key matrix $\mathbf{K}$. The scaling factor $\sqrt{d_s}$ is used to avoid large weights, especially when the dimensionality is high.

For an attention head $i$ in layer $l$, all inputs to the scaled dot-product attention come from layer $l-1$'s output. This implies a self-attention mechanism. The query, key, and value matrices are linear projection of $\mathbf{H}^{(l-1)}$. The head is defined as:
\begin{equation} \label{equ:head_attention}
\begin{aligned}
\text{head}_{i}^{(l)} &= \text{Attention}\big(\mathbf{Q}^{(l,i)}, \mathbf{K}^{(l,i)}, \mathbf{V}^{(l,i)}\big) \\
\text{where } &
\mathbf{Q}^{(l,i)} = \mathbf{H}^{(l-1)} \mathbf{W}^{(l,i)}_Q \\
& \mathbf{K}^{(l,i)} = \mathbf{H}^{(l-1)} \mathbf{W}^{(l,i)}_K \\
& \mathbf{V}^{(l,i)} = \mathbf{H}^{(l-1)} \mathbf{W}^{(l,i)}_V
\end{aligned}
\end{equation}
In the above equation, $\mathbf{W}^{(l,i)}_Q, \mathbf{W}^{(l,i)}_K, \mathbf{W}^{(l,i)}_V \in \mathbb{R}^{d \times d_s}$ are the corresponding matrices that project input $\mathbf{H}^{(l-1)}$ into the latent space of query, key, and value. Row $\tau$ of $\text{head}_i^{(l)}$ corresponds to an intermediate representation of a user's behavior sequence at timestamp $\tau$. 

Items in a user behavior sequence are produced one by one. The model should only consider previous items when predicting the next item. However, the aggregated value in Equation (\ref{equ:head_attention}) contains information of subsequent items, which makes the model ill-defined. Therefore we remove all links between row $\tau$ in $\mathbf{Q}$ and row $\tau'$ in $\mathbf{V}$ if $\tau > \tau'$.

After all attention heads are computed in layer $l$, their outputs are concatenate and projected by $\mathbf{W}^{(l)}_O \in \mathbb{R}^{d \times d}$, resulting in the final output $\mathbf{A}^{(l)} \in  \mathbb{R}^{m \times d}$ of a multi-head attention sub-layer:
\begin{equation}
\mathbf{A}^{(l)} = \text{Concat}\big(\text {head}_{1}^{(l)}, \text {head}_{2}^{(l)}, \cdots, \text{head}_{r}^{(l)}\big) \mathbf{W}^{(l)}_O
\end{equation}

{\bfseries Feed Forward Network}. Although previous items' information can be aggregated in the multi-head self-attention, it's still a linear model. To improve the representation power of our model, we apply a two-layer feed forward network to each position:
\begin{equation}
\label{equ:FFN}
\mathbf{F}^{(l)} = f\big(\mathbf{A}^{(l)} \mathbf{W}_{\text{FFN}}^{(l,1)} + \mathbf{b}_{\text{FFN}}^{(l,1)}\big) \mathbf{W}_{\text{FFN}}^{(l,2)} + \mathbf{b}_{\text{FFN}}^{(l,2)}
\end{equation}
where $\boldsymbol{W}_{\text{FFN}}^{ (k,1)}, \boldsymbol{W}_{\text{FFN}}^{ (k,2)}$ are both $d \times d$ matrices and $\boldsymbol{b}_{\text{FFN}}^{ (k,1)},\boldsymbol{b}_{\text{FFN}}^{ (k,2)}$ are $d$ dimensional vectors. Moreover $f$ is an activation function and we choose Gaussian Error Linear Unit as in work \cite{DBLP:journals/corr/HendrycksG16}. Nonlinear transformation is applied to all positions independently, meaning that no information will be exchanged across positions in this sub-layer.

{\bfseries Add \& Norm Sub-layer}. Training a multi-layers network is difficult because the vanishing gradient problem may occur in back-propagation. Residual neural network \cite{DBLP:conf/cvpr/HeZRS16} has shown its effectiveness in solving this problem. The core idea behind the residual neural network is to propagate outputs of lower layers to higher layers simply by adding them. If lower layer outputs are useful, the model can skip through higher layers to get necessary information. Let $\mathbf{X}$ be the output from lower layer and $\mathbf{Y}$ be the output from higher layer. The residual layer (or add) layer is defined as:
\begin{equation}
\text{Add} (\boldsymbol{X}, \boldsymbol{Y}) = \boldsymbol{X} + \boldsymbol{Y}
\end{equation}

In addition, to stabilize and accelerate neural network training, we apply Layer Normalization \cite{DBLP:journals/corr/BaKH16} to the residual layer output. Assuming the input is a vector $\boldsymbol{z}$, the operation is defined as:
\begin{equation}
\text{LayerNorm} (\boldsymbol{z}) = \boldsymbol{\alpha} \odot \frac{\boldsymbol{z} -\mu}{\sqrt{\sigma^2 + \epsilon}} + \boldsymbol{\beta}
\end{equation}
where $\mu$ and $\sigma$ are the mean and variance of $\boldsymbol{z}$, $\boldsymbol{\alpha}$ and $\boldsymbol{\beta}$ are learned scaling and bias terms, and $\odot$ represents an element-wise product.

{\bfseries Personal Preference Embedding Output}. SocialTrans encapsulates a user's previous behavior sequence into a $d$ dimensional embedding. Let $l_T$ be the number of layers in Transformer. The output of the $l_T$-th layer is $\mathbf{H}^{(l_T)}$. We take $\mathbf{h}^{(l_T)}_{m-1}$ as the user's personal preference embedding, where $\mathbf{h}^{(l_T)}_{m-1}$ is the last row of $\mathbf{H}^{(l_T)}$. The personal preference embedding is expressive because $\mathbf{h}^{(l_T)}_{m-1}$ aggregated all previous items information in multi-head self-attention layer. Moreover, stacking multiple layers provides personal preference embedding with the highly non-linearly expressive power.

\subsection{Socially Influenced Preference Modeling} \label{subsection:social_influence_preference}
\textit{Birds of a feather flock together}. A user’s behavior is influenced by his friends \cite{marsden1993network,mcpherson2001birds}. We should incorporate the social information to further model user latent factors. Meanwhile, different social connections or friends have different influence on a user. In other words, the learning of social-space user latent factors should consider different strengths in social relations. Therefore, we introduce a multi-head graph attention network (GAT) \cite{velivckovic2017graph} to select friends that are representative in characterizing users' socially influenced preferences. We also consider edge attributes to learn the context-dependent social influence. Next, we describe the input, output and our modified GAT in detail.

{\bfseries Input Embedding}. In §\ref{subsection:self_preference}, we described how to obtain a user’s personal preference embedding given his behavior sequence. Suppose we want to generate the socially influenced preference embedding of a user $u$. This module's inputs are personal preference embeddings of $u$ and his friends. Specifically, for a user $u$, the input of this module is $\{\hat{\mathbf{h}}^{(0)}_{u'} | u' \in \{u\} \cup N(u)\}$, where $\hat{\mathbf{h}}^{(0)}_{u} = \mathbf{h}^{(l_T)}_{m-1}$ and $N(u)$ is the set of user $u$'s friends.

\begin{figure}[h]
	\centering
	\includegraphics[width=.6\linewidth]{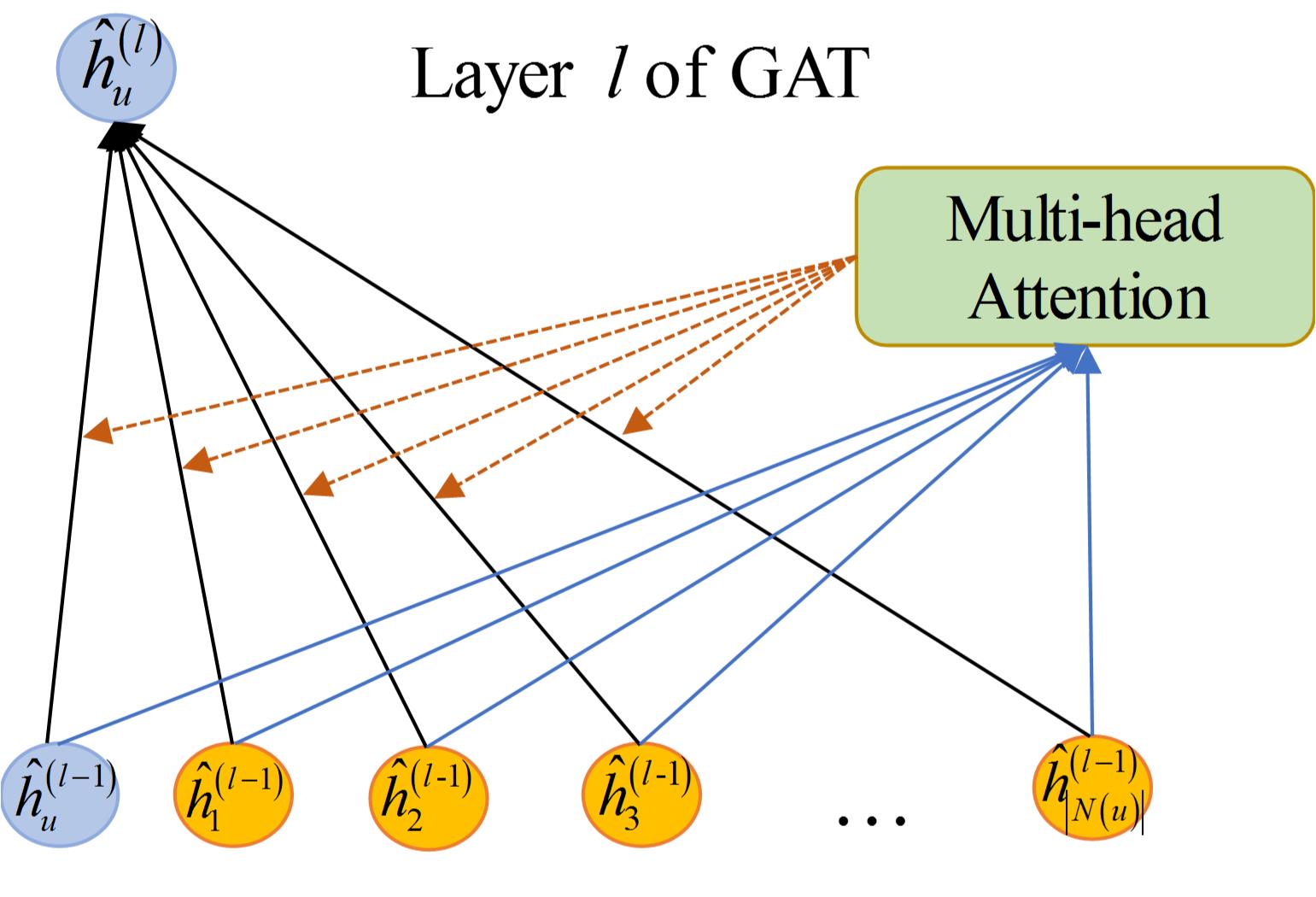}
	\caption{Socially influenced preference is modeled by a multi-head and context dependent GAT .}
	\label{figure:GAT}
\end{figure}

{\bfseries Graph Attention Network}. Veličković et al. \cite{velivckovic2017graph} introduces graph attention networks (GAT) which specifies different weights to different nodes in the neighborhood. Social influence is often context-dependent. It depends on both friends' preferences and the degree of closeness with friends. We use the GAT to aggregate contextual information from the user's friends. In this work, we propose an extended GAT that uses edge attributes and the multi-heads mechanisms. Figure \ref{figure:GAT} shows our modified version of GAT.

We first calculate the similarity score $\delta_{u,u'}^{(l)}$ between the target user's embedding $\hat{\mathbf{h}}^{(l-1)}_{u}$ and all of his neighbors' embedding $\hat{\mathbf{h}}^{(l-1)}_{u'}$:
\begin{equation} 
\label{equ:GAT_similarity}
\delta_{u,u'}^{(l)} = \big(\hat{\mathbf{W}}_Q^{(l)} \hat{\mathbf{h}}^{(l-1)}_{u}\big)^T \big(\hat{\mathbf{W}}_K^{(l)} \hat{\mathbf{h}}^{(l-1)}_{u'} \big) + \big(\hat{\mathbf{w}}_E^{(l)}\big)^T \mathbf{e}_{u,u'}
\end{equation}

Then we normalized the similarity score to a probability distribution:
\begin{equation}
\label{equ:GAT_norm_sim}
\kappa_{u,u'}^{(k)} = \frac{\exp(\delta_{u,u'}^{(k)})}{ \sum_{i \in N(u) \cup \{u\}} \exp(\delta_{u,i}^{(k)}) }
\end{equation}
where $\hat{\mathbf{W}}_Q^{(l)}, \hat{\mathbf{W}}_K^{(l)} \in \mathbf{R}^{d \times d}$ in Equation (\ref{equ:GAT_similarity}) are the query and key projection matrices similar to Equation (\ref{equ:head_attention}) in Transformer. $\mathbf{e}_{u,u'}$ is a vector of attributes corresponding to the edge between $u$ and $u'$, and $\hat{\mathbf{w}}_E^{(l)}$ is a weighted vector applied to these attributes. Equation (\ref{equ:GAT_similarity}) computes similarity score based on user's and friend's representation and their corresponding attributes. This enables us to combine both the preferences of friends and the degree of closeness with friends.

Intuitively, $\kappa_{u,u'}^{(l)}$ is the social influence strength of a friend $u'$ on the user $u$. We aggregate social influence of all friends as:
\begin{equation}
\label{equ:GAT_aggregate}
\hat{\mathbf{h}}^{(l)}_{u} = f\big(\sum_{i \in N(u) \cup \{u\}} \kappa_{u,i}^{(l)} \hat{\mathbf{W}}_V^{(l)} \hat{\mathbf{h}}^{(l-1)}_{i} \big)
\end{equation}
where $\hat{\mathbf{W}}_V^{(l)} \in \mathbf{R}^{d \times d}$ is value projection matrix and $f$ is a Gaussian Error Linear Unit as in Equation (\ref{equ:FFN}).

In practice, we find that the multi-head attention mechanism is useful to jointly capture context semantic at different subspaces. We extend our previous Equations (\ref{equ:GAT_similarity}), (\ref{equ:GAT_norm_sim}), and (\ref{equ:GAT_aggregate}) to:
\begin{equation}
\delta_{u,u'}^{(l,i)} = \big(\hat{\mathbf{W}}_Q^{(l,i)} \hat{\mathbf{h}}^{(l-1)}_{u}\big)^T \big(\hat{\mathbf{W}}_K^{(l,i)} \hat{\mathbf{h}}^{(l-1)}_{u'}\big) + \big(\hat{\mathbf{w}}_E^{(l,i)}\big)^T \mathbf{e}_{u,u'}
\end{equation}
\begin{equation}
\kappa_{u,u'}^{(l,i)} = \frac{\exp(\delta_{u,u'}^{(l,i)})}{ \sum_{j \in N(u) \cup \{u\}} \exp(\delta_{u,j}^{(l,i)}) }
\end{equation}
\begin{equation}
\hat{\mathbf{h}}^{(l,i)}_{u} = f\big(\sum_{j \in N(u) \cup \{u\}} \kappa_{u,j}^{(l,i)} \hat{\mathbf{W}}_V^{(l,i)} \hat{\mathbf{h}}^{(l-1)}_{j} \big)
\end{equation}
where $\hat{\mathbf{W}}_Q^{(l,i)}, \hat{\mathbf{W}}_K^{(l,i)}, \hat{\mathbf{W}}_V^{(l,i)} \in \mathbb{R}^{d_s \times d}$ and $\hat{\mathbf{w}}_E^{(l,i)}$ are corresponding parameters for head $i$. Finally, all $r$ heads are stacked and projected by $\hat{\mathbf{W}}^{(l)}_O \in \mathbf{R}^{d \times d}$  to get the final output embedding in layer $l$:
\begin{equation}
\hat{\mathbf{h}}^{(l)}_{u} = \hat{\mathbf{W}}^{(l)}_O \text [\hat{\mathbf{h}}^{(l,1)}_{u}; \hat{\mathbf{h}}^{(l,2)}_{u}; \cdots; \hat{\mathbf{h}}^{(l,r)}_{u}]
\end{equation}

{\bfseries Social Embedding Output}. We stack $l_G$ layers of GAT and take the final representation $\hat{\mathbf{h}}^{(l_G)}_{u}$ as the user $u$ socially influenced preference embedding. It encapsulates social information from user $u$ and his friends. Note that stacking too many layers of GAT may harm the model performance, because our analysis result shown in Figure \ref{figure:introduction_analysis} suggests that most social information come from one-hop friends. Thus, we use at most two layers of GAT.

\subsection{Rating \& Prediction} \label{subsection:recommendation}
A user's decisions depend on the dynamic personal preference and socially influenced preference from his friends. The final embedding representation is obtained by merging personal preference embedding and socially influenced preference embedding, which is defined as:
\begin{equation}
\tilde{\mathbf{h}}_u = \mathbf{W}_F [ \mathbf{h}^{(l_T)}_{m-1} ; \hat{\mathbf{h}}^{(l_G)}_{u}]
\end{equation}
where $\mathbf{W}_F \in \mathbf{R}^{d \times 2d}$ and $\tilde{\mathbf{h}}_u$ is user $u$'s final representation.

The probability that the next item will be $v$ is computed using a softmax function:
\begin{equation}
\label{equ:rating_softmax}
p(v_m = v | \hat{\mathbb{S}}_{t-1}^u) = \frac
{ \exp(\tilde{\mathbf{h}}_u^T \mathbf{w}_v) }
{ \sum_{i \in V} \exp(\tilde{\mathbf{h}}_u^T \mathbf{w}_i) }
\end{equation}

We train the model parameters by maximizing the log-probability of all observed sequences:
\begin{equation}
\sum_{u \in U} \sum_{t} \log p(v_m = v| \hat{\mathbb{S}}_{t-1}^u)
\end{equation}

To predict which item user $u$ will click, we compute all items' scores according to Equation (\ref{equ:rating_softmax}) and return a list of items with top $K$ scores for the recommendation. To avoid large computational cost in a real-world application, the approximate nearest neighbor search method \cite{DBLP:journals/cacm/AndoniI08} is used.

\section{Large Scale Implementation} \label{section:large_scale}
A real-world application may contain billions of users and millions of items. The main challenge to deploy the above model into online service is the large number of graph edges. For example, the social network of WeChat contains hundreds of billions of edges. The data is of multi-terabyte size and can not be loaded into the memory of any commercial server. Many graph operations may fail in this scenario. In addition, directly computing matching score function between user $u$'s representation vector $ \tilde{\mathbf{h}}_u $ and item representation matrix $\mathbf{W}$ is computationally costly. In this part, we discuss several implementation details about how we apply SocialTrans to large scale recommendation systems in industries.

{\bfseries Graph sampling}. Directly applying graph attention operation over the whole graph is impossible. A node can have thousands of neighbors in our data. We use a graph sampling technique to create a sub-graph containing nodes and their neighbors, which is computationally possible in a minibatch. Each node samples its $n$-hop sub-graph independently. Neighbors in each hop can be sampled simply by uniform sampling or can be sampled according to a specific edge attribute. For example, sampling by the number of commonly clicked items. It means a neighbor has a greater chance to be sampled if he clicks a greater number of items that are both clicked by a user and the neighbor. The sampling process is repeated several times for each node and implemented in a MapReduce style data pre-processing program. In this way, we can remove the graph storage component and reduce communication costs during the training stage.

{\bfseries Sampling negative items}. There are millions of candidate items in our settings. Many methods require computing matching score function between $\tilde{\mathbf{h}}_u$ and $\mathbf{w}_v$. After that, the softmax function is applied to obtain the predicted probability. We use negative sampling to reduce the computational cost of the softmax function. In each training minibatch, we sample a set of 1000 negative items $J$ shared by all users. The probability of next item will be $v$ in this minibatch is approximated as:
\begin{equation}
p(v_m = v| \hat{\mathbb{S}}_{t-1}^u) = \frac
{ \exp(\tilde{\mathbf{h}}_u^T \mathbf{w}_v) }
{ \sum_{i \in \{v\} \cup J} \exp(\tilde{\mathbf{h}}_u^T \mathbf{w}_i) }
\end{equation}
The negative item sampling probability is proportional to its appearance count in the training data set. Using negative sampling technique provides an approximation of the original softmax function. Empirically, we do not observe a decline in performance.

We use Adam \cite{DBLP:journals/corr/KingmaB14} for optimization because of its effectiveness. Directly applying Adam will be computational infeasible because its update is applied to all trainable variables. The items' representation matrix $\mathbf{W}$ will be updated although many items do not appear in that minibatch. Here we adopt a slightly modified version of Adam, which only updates items appeared in the minibatch and other trainable variables. We update each trainable variable $\theta$ at the training step $k$ according to the following equations:

\begin{equation} \label{equ:adam}
\theta_k = 
\begin{cases}
\theta_{k-1} - \eta \frac{ m_k / (1-\beta_1^k) } {\sqrt{n_k / (1-\beta_2^k)} + \epsilon} & \theta \in \Theta_k \\
\theta_{k-1} & \text{otherwise}
\end{cases}
\end{equation}

\begin{equation} \label{equ:adam_m}
m_k = 
\begin{cases}
\beta_1 m_{k-1} + (1-\beta_1) \nabla_{\theta} L(\theta_{k-1}) & \theta \in \Theta_k \\
m_{k-1} & \text{otherwise}
\end{cases}
\end{equation}

\begin{equation} \label{equ:adam_v}
n_k = 
\begin{cases}
\beta_2 n_{k-1} + (1-\beta_2) (\nabla_{\theta} L (\theta_{k-1}))^2 & \theta \in \Theta_k \\
n_{k-1} & \text{otherwise}
\end{cases}
\end{equation}
In the above equations, $\beta_1,\beta_2,\epsilon$ are Adam's hyperparameters. We fix them to 0.9, 0.999 and 1e-8 respectively. $\Theta_k$ are sets of items representation parameters and other trainable variables appeared in the minibatch $k$. For items not appeared, their parameters and statistics in Adam remain unchanged at this step.
Empirically, combining negative sampling and sparse Adam update techniques can speed up the training procedure 3-5x times when there are millions of items.

{\bfseries Multi-GPU training}. The socially influenced preference module's inputs are personal preference embeddings of a user and his friends. These personal preference embeddings are computationally expensive. It's necessary to utilize multiple GPUs on a single machine to speed up the training procedure. We adopt the data parallelism paradigm to utilize multiple GPUs. Each minibatch is split into sub-mini batches with equal sizes. And each GPU runs the forward and backward propagation over one sub-minibatch using the same parameters. After all backward propagation is done, gradients are aggregated and we use the aggregated gradients to perform parameters update. For training efficiency, we train our model with a large minibatch size until the memory can not hold more samples.

{\bfseries Embedding Generation}.
Since the size of input and output data is large, generating fusion embedding results on a single machine requires a large amount of disk space. Since the generation procedure is less computationally expensive, we implement it on a cluster without GPUs. We split the generation procedure into three stages: 
\begin{enumerate}
	\item The first stage is generating personal preference embedding $\mathbf{h}^{(l_T)}_{m-1}$ for all users and item embedding matrix $\mathbf{W}$. This stage is less computationally expensive than the training stage. And we implement it on a distributed Spark \cite{DBLP:conf/hotcloud/ZahariaCFSS10} cluster.
	\item The second stage is to retrieve the user's and their friends' personal preference embedding. This stage requires lots of disk access and network communication. It is implemented by a SparkSQL query on a distributed data warehouse.
	\item The final stage is generating the users' social influenced preference embedding $\hat{\mathbf{h}}^{(l_G)}_{u}$ and fusing it with users' personal preference embedding $\mathbf{h}^{(l_T)}_{m-1}$ to get the final embedding $\tilde{\mathbf{h}}_u$. Another Spark program is implemented to generate the final user embedding.
\end{enumerate}  
The intermediate results and all embedding outputs are stored in a distributed data warehouse. Downstream tasks retrieve the results to provide online service.

\section{EXPERIMENTS} \label{section:experiments}
In this section, we first describe experimental data sets, compared methods, and evaluation metrics. Then, we show results of offline experiments on two data sets and online valuation of our method on an article recommendation system in WeChat. Specifically, we aim to answer the following questions:

\textbf{Q1:} How does SocialTrans outperform the state-of-the-art methods for the recommendation tasks?

\textbf{Q2:} How does the performance of SocialTrans change under different circumstances?

\textbf{Q3:} What is the quality of generated user representation for online services?

\subsection{Data Sets}
We evaluate our model in three data sets. For offline evaluation, we tested on a benchmark data set \emph{Yelp} and a data set \emph{WeChat Official Accounts}. For online evaluation, we conducted experiments on \emph{WeChat Top Stories}, a major article recommendation application in China\footnote{All users in data sets of WeChat are anonymous.}. The statistics of those data sets are summarized in Table \ref{table:data_statistics}. We describe the detailed information as follows:

\emph{Yelp}\footnote{https://www.yelp.com/dataset}. Yelp is a popular review website in the United States. Users can review local businesses including restaurants and shops. We treat each review from 01/01/2012 to 11/14/2018 as an interaction. This data set includes 3.3 million user reviews, more than 170,000 businesses, more than 250,000 users and 4.7 million friendship links. The last 150 days of reviews are used as a test set and the training set is constructed from the remaining days. Items that do not exist in the training set are removed from the test set. For each recommendation, we use the user's own interaction and friends' interactions before the recommendation time.

\emph{WeChat Official Accounts}. WeChat is a Chinese messaging mobile app with more than one billion active users. WeChat users can register an official account, which can push articles to subscribed users. We sample users and their social networks restricted to an anonymous city in China. We treat each official account reading activity as an item click event. The training set is constructed from users' reading logs in June 2019. The first four days of July 2019 are kept for testing. We remove items appeared less than five times in training data to ensure the quality of recommendation. After processing, this data set contains more than 736,000 users, 48,000 items and each user has an average of 81 friends.

\emph{WeChat Top Stories}. WeChat users can receive articles recommendation service in Top Stories, whose contents are provided by WeChat official accounts. Here we treat each official account reading activity as an item click event. Training logs are constructed from users' reading logs in June 2019. Testing is conducted on an online environment in five consecutive days of July 2019. This data set contains billions of users and millions of items.  In this data set, we keep the specific values for business secret.

%\begin{savenotes}
\begin{table}[h]
	\centering
	\begin{tabular}{lccc}
		\hline

		& Yelp & \begin{tabular}[c]{@{}c@{}}WeChat\\ Official Accounts\end{tabular} & \begin{tabular}[c]{@{}c@{}}WeChat\\ Top Stories\end{tabular} \\ \hline
		Users & 251,181 & 736,984 & $\sim$ Billions \\
		Items & 173,567 & 48,150 & $\sim$ Millions \\
		Events & 3,260,808 & 11,053,791 & $\sim$ Tens of Bil. \\
		Relations & 4,753,048 & 59,809,526 &  - \\
		Avg. friends & 18.92 & 81.15 & $\sim$ Hundreds \\
		Avg. events & 12.98 & 14.99 & $\sim$ Tens \\
		Start Date & 01/01/2012 & 01/06/2019 & 01/06/2019 \\
		End Date & 11/14/2018 & 04/07/2019 & 31/06/2019 \\
		Evaluation & Offline & Offline & Online \\ \hline
	\end{tabular}
    \caption{The statistics of the experimental data sets.}
	\label{table:data_statistics}
\end{table}
%\end{savenotes}

\subsection{Offline Model Comparison - Q1}
In this subsection, we evaluate the performance of different models on \emph{Yelp} and \emph{WeChat Official Accounts} data sets.

\subsubsection{Evaluation Metrics.} In the offline evaluation, we are interested in recommending items directly. Each algorithm recommends items according to their ranking scores. We evaluate two widely used ranking-based metrics: Recall@K and Normalized Discounted Cumulative Gain (NDCG).

\emph{Recall@K} measures the average proportion of the top-K recommended items that are in the test set.

\emph{NDCG} measures the rank of each clicked user-item click pair in a model. It's formulated as $NDCG = \frac{1}{\log_2{ (1 + \text{rank})}}$. We report the average value over all the testing data for model comparison.

\subsubsection{Compared Models.} We compare SocialTrans with several state-of-the-art baselines. The details of these models are described as follows:

\textbf{POP}: a rule-based method, which recommends a static list of popular items. The rank of each item is sorted according to the number of appearances in the training data.

\textbf{GRU4Rec} \cite{hidasi2015session}: a sequence-based approach that captures users' personal preferences by a recurrent neural network. Items are recommended according to these personal preferences.

\textbf{SASRec} \cite{DBLP:conf/icdm/KangM18}: another sequence-based approach that uses a multi-layer Transformer \cite{DBLP:conf/nips/VaswaniSPUJGKP17} to capture the users' personal preferences evolved over time.

\textbf{metapath2vec} \cite{DBLP:conf/kdd/DongCS17}: it is an unsupervised graph-based approach. It first generates meta-path guided random walk sequences. We utilize the user-item bipartite graph and let the meta-path to be "user-item-user". Embedding representation of each user/item is learned by using these sequences to preserve neighboring proximity. For recommendation, the rank score is computed by the dot product of user and item embedding.

\textbf{DGRec} \cite{song2019session}: an approach utilizing both temporal and social factors. In this approach, each user representation is generated by a fusion of personal and socially influenced preference. A recurrent neural network is used to model a user's short term interest. And a unique user embedding is learned to capture the long term interest. Socially influenced preference is captured by a graph attention neural network without considering edge attributes and the multi-head attention mechanism.

\textbf{Social-GAT}: our modified version of graph attention neural network (GAT) \cite{velivckovic2017graph}. Here we represent a user's personal preference embedding as an average embedding of his previously clicked items. Then GAT over social graph is applied to capture socially influenced preference. A user's final representation is generated by fusing personal preference and socially influenced preference.

\subsubsection{Evaluation details.} We train all models with a batch size of 128 and a fixed learning rate of 0.001. We use our modified version of Adam \cite{DBLP:journals/corr/KingmaB14} for optimization (mentioned in §\ref{section:large_scale}). For all models, The dimension of each user and item are fixed to 100. We cross-validated other model parameters using 80\% training logs and leave out 20\% for validation. To avoid overfitting, the dropout technique \cite{DBLP:journals/jmlr/SrivastavaHKSS14} with rate 0.1 is used. The neighborhood sampling size is empirically set from 20 to 50 in GAT layers.

\subsubsection{Comparative Results} We summarize the experimental results of SocialTrans and baseline models in Table \ref{table:data_offline_comparison}. We have the following findings:
\begin{itemize}
	\item SocialTrans outperforms other baseline models. It achieves a 9.56\% relative recall@20 improvement compare to Social-GAT in the Yelp data set. For Wechat Official Accounts, the relative improvement of recall@10 is 9.62\%.
	\item Sequence-based methods GRU4Rec and SASRec perform better than POP and metapath2vec. Because the last two methods do not consider the factor that user interests will evolve over time.
	\item DGRec does not get a greater performance boost than sequence-based methods on both data set. DGRec uses a unique user embedding to capture the user's long term interest. This can lead to a large increase in model parameters because the number of users are large in both data sets. Similar results are observed in metapath2vec, which only uses a unique embedding to represent a user. We believe that DGRec and metapath2vec are suffered from the problem of overfitting due to the large number of model parameters in these data sets.
	\item SocialTrans consists of two components - user's personal preference and socially influenced preference. If we only consider the preferences of users themselves, SocialTrans degenerates to SASRec. Without the consideration of users' preference being dynamic, SocialTrans degrades to Social-GAT. Notice that SocialTrans, SASRec, and Social-GAT provide a strong performance, which shows the effectiveness of our model's different variations.
	\item SocialTrans and Social-GAT achieve a greater performance gain than other methods in these data sets, that means social factor has great potential business value. We believe the performance boost comes from special services provided by applications. In the WeChat Official Accounts scenario, users can share official accounts with their friends. This implies that official accounts that are subscribed by more friends are more likely to be recommended and clicked.
\end{itemize}
In summary, the comparison results show that (1) both temporal and social factors are helpful for recommendations; (2) restricted model capacity is critical to avoid overfitting in real-world data sets; (3) our model outperforms baseline methods.

\begin{table}[h]
	\centering
	\begin{tabular}{lllcc}
		\hline
		& \multicolumn{2}{c}{Yelp} & \multicolumn{2}{c}{WeChat Official Acct.}             \\
		Model        & recall@20    & NDCG      & \multicolumn{1}{l}{recall@10} & \multicolumn{1}{l}{NDCG} \\ \hline
		POP          &   1.05\%       &  9.39\%    &  3.87\% & 12.22\% \\
		GRU4Rec \cite{hidasi2015session}      & 5.84\%       & 11.68\%   & 6.27\%                        & 13.03\%                  \\
		SASRec \cite{DBLP:conf/icdm/KangM18}       & 6.18\%       & 12.83\%   & 7.01\%                        & 13.50\%                  \\
		metapath2vec \cite{DBLP:conf/kdd/DongCS17} & 1.06\%       & 9.41\%    & 5.23\%                        & 12.42\%                  \\
		DGRec \cite{song2019session}        & 5.92\%       & 12.71\%   & 6.59\%                        & 13.04\%                  \\
		Social-GAT   & 6.27\%       & 12.92\%   & 8.52\%                        & 14.45\%                  \\
		SocialTrans  & \textbf{6.87\%} & \textbf{13.23\%}   & \textbf{9.75\%} & \textbf{15.19\%}  \\ \hline
	\end{tabular}
	\caption{Offline comparison of different models.}
	\label{table:data_offline_comparison}
\end{table}

\subsection{Offline Study - Q2}
We are interested in the performance of SocialTrans under different circumstances. First, we analyze the performance of different models for users with different number of friends. Then, we analyze the performance of SocialTrans with different number of layers.

\subsubsection{Number of Friends.}
In social recommendation systems, different users have a different number of friends. When a user has very few friends, the number of items clicked by his friends is limited and the potential of utilizing social information could be limited. On the other hand, when a user has many friends, there are many items clicked by his friends both in the past and in the future. In this case, SocialTrans could leverage social information, learn more, and make better predictions. 

\begin{figure}[h]
	\centering
	\includegraphics[width=\linewidth]{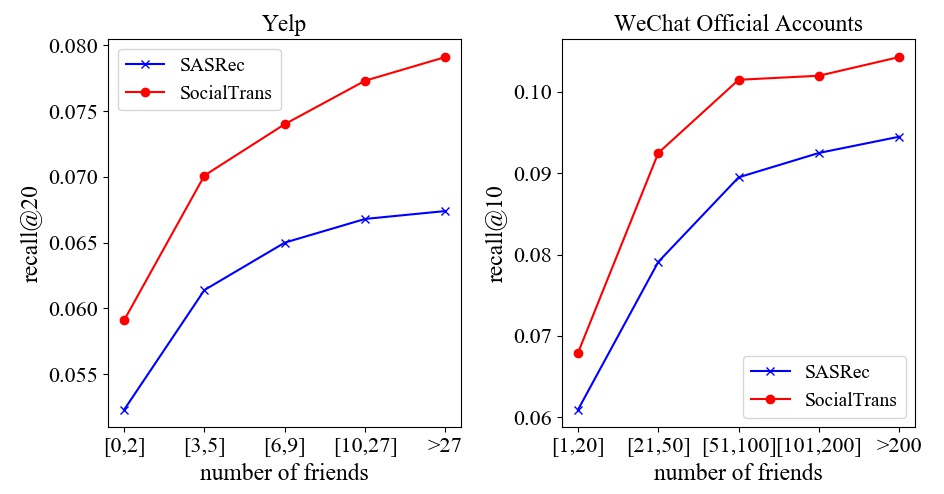}
	\caption{User performance with different friends number.}
	\label{figure:offline_evaluation_friends_group}
\end{figure}

We investigate the recommendation performance of SocialTrans and SASRec \cite{DBLP:conf/icdm/KangM18} on users with a different number of friends. Figure \ref{figure:offline_evaluation_friends_group} shows the result. It can be seen that SocialTrans consistently outperforms SASRec in all groups. In the Yelp data set, the improvement of SocialTrans over SASRec becomes larger when users have more friends, which matches our analysis in the previous paragraph. The relative improvement of SocialTrans is 5.35\% for group 0-2 and is 16.15\% for group $>$27. In the WeChat Official Accounts data set, the best relative improvement is achieved at group 21-50.

\subsubsection{Number of Transformer and GAT Layers.}
The performance of deep learning methods can be empirically improved when stacking more layers. We are interested in how the performance of SocialTrans change with different layers of Transformer or GAT. Table \ref{table:data_offline_layers} summarizes the result. Stacking more Transformer layers can largely boost performance. In the WeChat Official Account data set, the recall@10 metric improves relatively by 7.45\% when the number of Transformer layers increases from 1 to 3. On the other hand, stacking more GAT layers can improve the result, but the improvement is not as significant as stacking more Transformer layers. The reason behind this is that social influence decays very fast and most of the information is provides by one-hop neighbors, which matches the analysis result in Figure \ref{figure:introduction_analysis}.

\begin{table}[h]
	\begin{tabular}{llllll}
		\hline
		&                                                      & \multicolumn{2}{c}{Yelp} & \multicolumn{2}{c}{WeChat Official Acct.} \\
		\begin{tabular}[c]{@{}l@{}}Trans. \\ layers\end{tabular} & \begin{tabular}[c]{@{}l@{}}GAT \\ layers\end{tabular} & recall@20     & NDCG     & recall@10                                   & NDCG                                  \\ \hline
		1 & 1 & 6.32\% & 12.87\% & 8.99\% & 14.88\% \\
		2 & 1 & 6.62\% & 13.09\% & 9.37\% & 14.97\% \\
		3 & 1 & 6.67\% & 13.11\% & 9.66\% & 15.12\% \\
		3 & 2 & 6.87\% & 13.23\% & 9.75\% & 15.19\% \\ \hline
	\end{tabular}
	\caption{SocialTrans performance with different layers.}
	\label{table:data_offline_layers}
\end{table}

\subsection{Online Evaluation - Q3} \label{subsection:online_evaluation}
\subsubsection{Evaluation Procedure.} 
\begin{figure}[h]
	\centering
	\includegraphics[width=.6\linewidth]{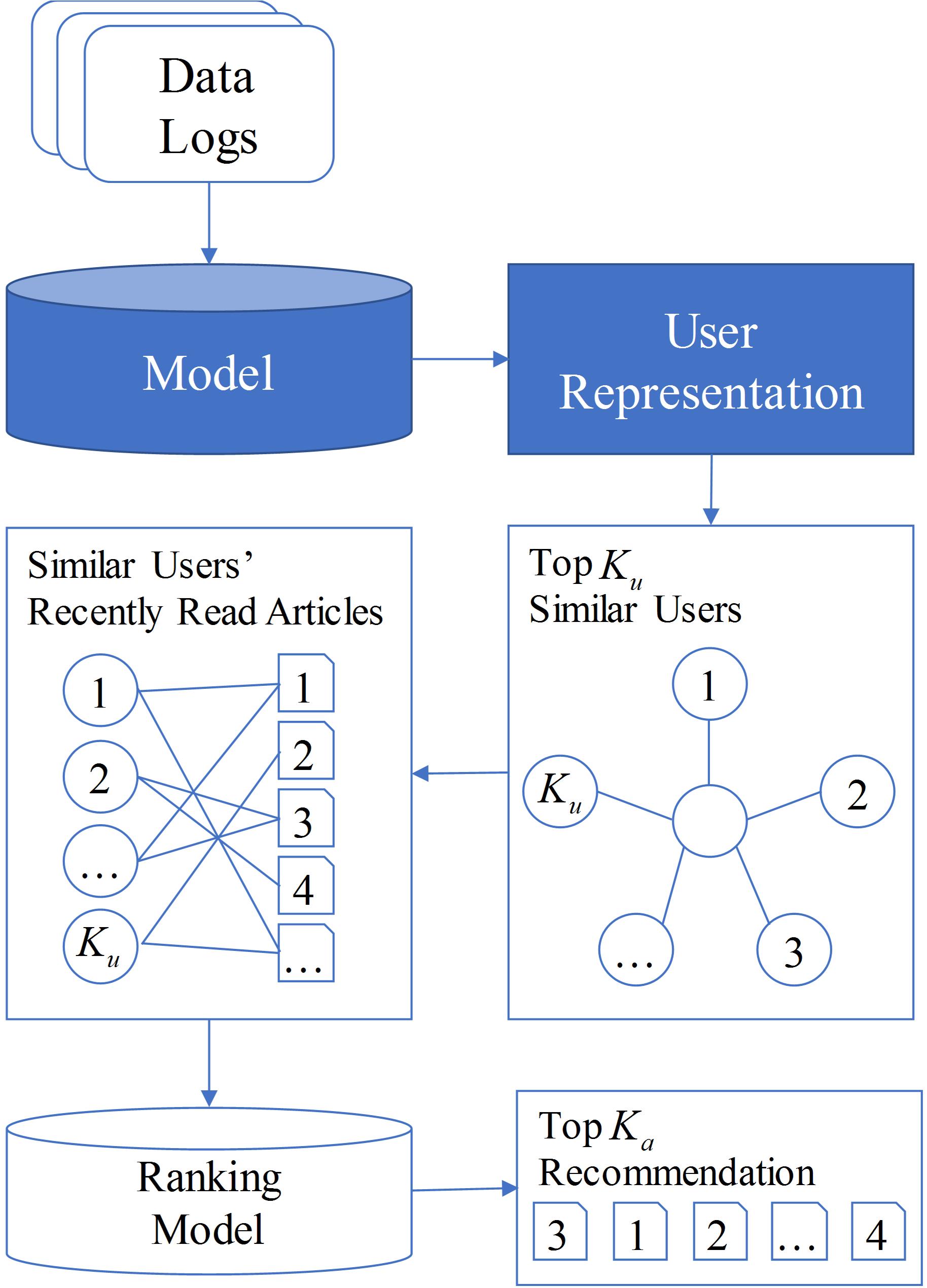}
	\caption{Online evaluation procedure for one model. Here we highlight model component and its generated user representation. Only these components are changed and others remain unchanged. }
	\label{figure:online_evaluation_procedure}
\end{figure}

To verify the effectiveness of our model, we establish an online A/B testing between SocialTrans and competition models in WeChat Top Stories, a major article recommendation engine in China. We divide online traffics into equal-sized portions and each model influences only one portion. Recommending $K_a$ items to users directly is difficult since it needs a great change in the online system. Furthermore, this direct approach wastes lots of computational resources because new articles are created very quickly. In this scenario, we use user-based collaborate filtering method shown in Figure \ref{figure:online_evaluation_procedure} as an indirect evaluation approach.

This approach only changes the recall component in the online serving system and is less computationally expensive as compared to the direct approach. Each model first generates a fixed-size representation for each user. For SocialTrans, this representation is user's fusion embedding vector. After that, the user-pair similarity is computed using the above representation to find top $K_u$ similar users. Since it's impossible to compute the similarity of all user-user pairs, we adapt approximate nearest neighbor algorithm SimHash \cite{DBLP:journals/cacm/AndoniI08}. We choose $K_u$ as 300. After that, top $K_u$ similar users' recently read articles are fed into the ranking model whose parameters are fixed in our evaluation. Finally, a list of articles with top $K_a$ score is presented to the user. For safety reasons, this evaluation only uses a very small portions of the online traffic.

\subsubsection{Evaluation Metrics \& Compared Models}
In online services, we are interested in CTR (click-through rate), which is a common evaluation metric in recommendation systems. It is the number of clicks divided by the number of recommended articles. We train models with the logs of June 2019 and evaluate them on the online A/B testing platform in five consecutive days of July 2019. Here we compare our model with the following models:

\textbf{metapath2vec} \cite{DBLP:conf/kdd/DongCS17}: a graph-based approach that takes bipartite user-item graph as input and generates a unique user embedding for each user. We choose "user-item-user" as the meta-path and take their embedding as the representation of users.

\textbf{SASRec} \cite{DBLP:conf/icdm/KangM18}: a multi-layer Transformer \cite{DBLP:conf/nips/VaswaniSPUJGKP17} model. We take the last layer of Transformer as the representation of users.

\textbf{SocialTrans}: our proposed model with a three-layer Transformer and a two-layer GAT. We take the fusion embedding as the representation of users.

\subsubsection{Online Results}
\begin{figure}[h]
	\centering
	\includegraphics[width=\linewidth]{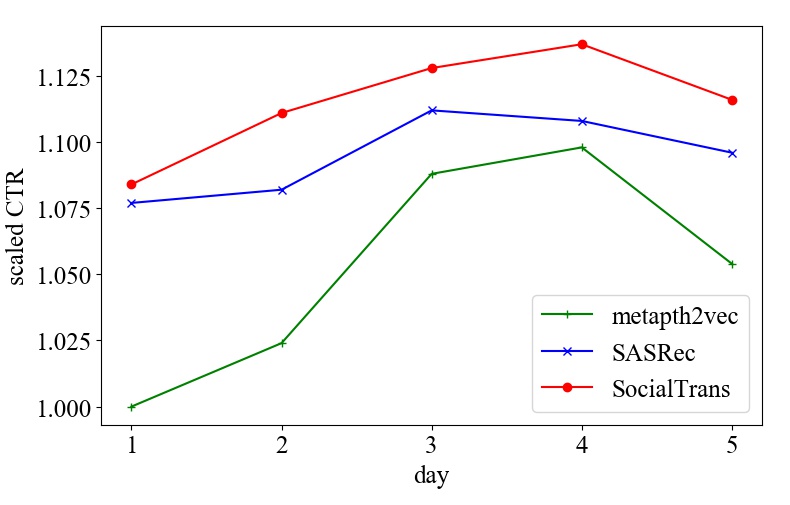}
	\caption{Online CTRs of different models in five days.}
	\label{figure:online_evaluation_details}
\end{figure}

Table \ref{table:data_online_comparison} and Figure \ref{figure:online_evaluation_details} show the result of SocialTrans and its competitors. We choose the CTR of metapath2vec on the first day as a base value. And we scale all CTRs by this base value. It can be observed that SocialTrans consistently outperforms metapath2vec and SASRec in five days, with an average of 5.98\% relative improvement compared to metapath2vec, which shows that the quality of user embedding in SocialTrans is better than baseline models.

Notice that the improvement in online evaluation is smaller that in the offline evaluation. This is because we chose an indirect evaluation approach - collaborative filtering, which requires fewer computational resources as compared to the direct recommendation approach when items are added or fading very fast. This provides an economical way to verify a new model.

\begin{table}[h]
	\begin{tabular}{lccc}
		\hline
		\multicolumn{1}{c}{} & \begin{tabular}[c]{@{}c@{}}Avgerage\\ Scaled CTR\end{tabular} & \begin{tabular}[c]{@{}c@{}}Relative\\ Improvement\end{tabular} \\ \hline
		metapath2vec \cite{DBLP:conf/kdd/DongCS17} & 105.3\% & 0\% \\
		SASRec \cite{DBLP:conf/icdm/KangM18} & 109.5\%  & +3.99\%  \\
		SocialTrans & 111.5\%  & +5.89\%      \\ \hline 
	\end{tabular}
	\caption{Online CTRs of different models in five days. We use the first day's CTR of metapath2vec as a base value to scale all CTRs.}
	\label{table:data_online_comparison}
\end{table}

\section{RELATED WORK} \label{section:realted_work}
\subsection{Sequential Recommendation}
Sequential recommendation algorithms use previous interactions to predict a user’s next interaction in the near future. Different from general recommendation, sequential recommendation always views the interactions as a sequence of items in time order. Previous sequential recommendation  methods have been based on Markov chains, these methods construct a transition matrix to solve sequential recommendation. Factorizing Personalized Markov Chains (FPMC) \cite{rendle2010factorizing} mixes Markov chains (MC) and matrix factorization (MF) together and estimates a transition cube to learn an transition matrix for each user. FPMC cannot catch relationships between items in a sequence, because it assumes each item independently affects the user's next interaction. HRM \cite{wang2015learning} could well capture both sequential behavior and users’ global interest by including transaction and user representations in prediction. Recently, there have been many methods that apply deep learning to recommendation systems emerged. Restricted Boltzmann Machine (RBM) \cite{salakhutdinov2007restricted} and Autoencoders \cite{sedhain2015autorec} have been the earliest methods, and RBM has been proved to be one of the best performing Collaborative Filtering models. Caser \cite{tang2018personalized} converts embedding of items in a sequence into a square matrix and uses CNN to learn user local features as a sequential pattern. However, CNN could hardly capture long-term dependencies between items and users' global features are always important for the recommendation. On the other hand, RNN has been widely used to model global sequential interactions \cite{yu2016dynamic}. GRU4Rec \cite{hidasi2015session} has been a representative method based on RNN for the sequential recommendation, which uses the final hidden state of GRU to delineate the user current preference. NARM and EDRec \cite{li2017neural,loyola2017modeling} use the attention mechanism to improve the effect of GRU4Rec. SASRec \cite{DBLP:conf/icdm/KangM18} use a multi-layer Transformer to capture user's dynamic interest evolved over time. SHAN \cite{ying2018sequential} proposed a two-layer hierarchical attention network to take both user-item and item-item interactions into account. These models assume that a user's behavior sequence is dynamically determined by his personal preference. However, they do not utilize social information, which is important in social recommendation tasks.

\subsection{Social Recommendation}
In online social networks, one common assumption is that a user’s preference is influenced by his friends. So introducing social relationships could improve the effectiveness of the model. Besides, for recommendation
systems, using friends' information of users could effectively alleviate cold start and data sparsity problems. There have been many studies that models the influence of friends on user interests from different aspects. Most proposed models are based on Gaussian or Poisson matrix factorization.
Ma et al. \cite{ma2011recommender} proposed a matrix factorization framework with social regularization such that the distances of connected users' embedding vectors are small. TrustMF \cite{yang2016social} adopts a matrix factorization technique to map users into low-dimensional potential feature spaces according to their trust relationship. SBPR \cite{zhao2014leveraging} is an approach that utilizes social information for training instance selection. Recently, many studies leveraged deep neural networks and network embedding approaches
to solve social recommendation problems because of their powerful performance. The NCF model \cite{he2017neural} leverages a multi-layer perceptron to learn the user-item interaction function. NSCR \cite{wang2017item} enhances the NCF model by plugging a pairwise pooling operation and extends NCF to cross-domain social recommendations.
Previous methods neglected the weight of the relationship edge, and different types of edges should play different roles. mTrust \cite{tang2012mtrust} and eTrust \cite{tang2012etrust} could model trust evolution, integrating multi-faceted trust relationships into traditional rating prediction algorithms in order to reliably evaluate their strengths. TBPR \cite{wang2016social} and PTPMF \cite{wang2017learning} distinguish between strong and weak ties of users for the recommendation in social networks. These approaches assume that a user's behavior sequence is determined by his personal preference and his socially influenced preference. However, these methods model users' personal preferences as static and social influences as context-independent.

\subsection{Graph Convolutional Networks}
Graph Convolutional Networks (GCN) has been a powerful technique to encode graph nodes into low-dimensional space and has been proven to have the ability to extract features from graph-structured data \cite{defferrard2016convolutional}. Kipf and Welling \cite{kipf2016semi} used GCNs for semi-supervised graph classification and achieved the state-the-of-art performance. GCNs combine the features of the current node and its neighbors, and handle edge features easily. However, in the original GCNs, all neighbors' weights are fixed when using convolution filters to update the node embeddings. GATs \cite{velivckovic2017graph} use an attention mechanism to assign different weights to different nodes in the neighborhood.  In the area of recommendation, PinSage \cite{ying2018graph} uses efficient random walks to structure the convolutions and is scalable to recommendations on large-scale networks. GCMC \cite{berg2017graph} proposes a graph auto-encoder framework based on differentiable message passing on the bipartite user-item graph and provides users' and items' embedding vectors for the recommendation. SocialGCN \cite{wu2018socialgcn} uses the ability of GCNs to catch how users’ interests are influenced by the social diffusion process in social networks. DGRec \cite{song2019session} proposes a method based on a dynamic-graph-attention neural network. DGRec uses RNN to model user short-term interest and uses GAT to model social influence between users and their friends. SocialTrans uses a multi-layers Transformer to model users' personal preference. We show by experiments that SocialTrans outperforms DGRec both on the Yelp data set and on the WeChat Official Accounts data set.

\section{CONCLUSION} \label{section:conclusion}
On social networks platforms, a user's behavior is based on his personal interests and is socially influenced by their friends. It is important to study how to consider both of these two factors for social recommendation tasks. In this paper, we presented SocialTrans, a model that jointly captures users' personal preference and socially influenced preference. SocialTrans uses Transformer and GAT, two state-of-the-art models, as building blocks. We conducted extensive experiments to demonstrate the superiority of SocialTrans over previously state-of-the-art models. On the offline data set Yelp, SocialTrans achieves 11.16\% \textasciitilde 17.63\% relative improvement of recall@20 as compared to sequence-based methods. On the WeChat Official Accounts data set, SocialTrans achieves a 39.08\% relative improvement over the state-of-the-art sequence-based method. In additional, we deployed and tested our model in WeChat Top Stories, a major article recommendation platform in China. Our online A/B testing show that SocialTrans achieves a 5.89\% relative improvement of click-through rate against metapath2vec. In the future, we plan to extend SocialTrans to take advantages of more side information of given recommendation tasks (e.g., other available attributes of users and items). Also, social networks are not static and they evolve over time. It would be interesting to explore how to deal with dynamic social networks.

%%
%% The next two lines define the bibliography style to be used, and
%% the bibliography file.
\bibliographystyle{ACM-Reference-Format}
\bibliography{main}

\end{document}